\documentclass[12pt]{ronbun}

\usepackage{amsmath,amssymb,graphicx,color}
\usepackage{cite}
\usepackage{bm}
\usepackage{dcolumn}

\newcommand{\nn}{\nonumber}
\newcommand{\e}{{\rm e}}

\newcommand{\al}{\alpha}

\numberwithin{equation}{section}

\begin{document}

\begin{flushright}
\parbox{4.2cm}
{KEK-TH-1021 \hfill \\
{\tt hep-th/0507029}
 }
\end{flushright}

\vspace*{1.1cm}

\begin{center}
 \Large\bf Thermodynamic Behavior of Fuzzy Membranes \\
in PP-Wave Matrix Model 
\end{center}
\vspace*{1.5cm}
\centerline{\large  Hyeonjoon Shin$^{\dagger a}$ and Kentaroh
Yoshida$^{\ast b}$}
\begin{center}
$^{\dagger}$\emph{ BK 21 Physics Research
    Division and Institute of Basic Science\\ 
    Sungkyunkwan University,
    Suwon 440-746, South Korea 
}\\ 
$^{\ast}$\emph{ Theory Division, High Energy Accelerator Research 
Organization (KEK),\\
Tsukuba, Ibaraki 305-0801, Japan.} 
\\
\vspace*{1cm}
 $^{a}$hshin@newton.skku.ac.kr \qquad $^{b}$kyoshida@post.kek.jp
\end{center}

\vspace*{1.5cm}

\centerline{\bf Abstract}
\vspace*{0.5cm}
We discuss a two-body interaction of membrane fuzzy spheres in a pp-wave
matrix model at finite temperature by considering a fuzzy sphere rotates
with a constant radius $r$ around the other one sitting at the origin in
the SO(6) symmetric space. This system of two fuzzy spheres is
supersymmetric at zero temperature and there is no interaction between
them. Once the system is coupled to the heat bath, supersymmetries are
completely broken and non-trivial interaction appears.  We numerically
show that the potential between fuzzy spheres is attractive and so the
rotating fuzzy sphere tends to fall into the origin. The analytic
formula of the free energy is also evaluated in the large $N$ limit. It
is well approximated by a polylog-function.

\vfill

\thispagestyle{empty}
\setcounter{page}{0}

\newpage 

\section{Introduction and Summary}

The basic degrees of freedom of string theory and M-theory are fully
encoded in matrix models \cite{BFSS,IKKT,DVV}. The matrix models are
believed to give non-perturbative formulations of string theory and
M-theory. In particular, the BFSS matrix model is supersymmetric matrix
quantum mechanics. It is believed to describe a discrete light-cone
quantization of M-theory. It also describes the low-energy dynamics of
$N$ D0-branes of type IIA superstring theory \cite{Witten}. In addition
it is a matrix regularization of the light-cone action for the
supermembrane in eleven dimensions \cite{dWHN}.

Matrix model thermodynamics is also interesting in relation to black hole
physics. One motivation for understanding the behavior of the BFSS
matrix model at finite temperature comes from the conjecture that their
finite temperature states are related to black hole states of type IIA
supergravity \cite{KS,BFKS}. This idea was studied in a series of papers
by Kabat, Lifschytz and Lowe \cite{KLL}. The BFSS matrix model at finite
temperature is also investigated in \cite{AMS}. Some features are
discussed in \cite{others} (For a review of brane thermodynamics, 
see \cite{Martinec:review}). 

By the way, a matrix model on a pp-wave background was proposed
by Berenstein-Maldacena-Nastase (BMN) \cite{BMN}, and it has been intensively
studied. The background of this matrix model is given by the maximally
supersymmetric pp-wave background \cite{KG}:
\begin{eqnarray}
\label{pp}
ds^2 &=& -2dx^+dx^- -\left(
\sum_{i=1}^3\left(\frac{\mu}{3}\right)^2 (x^i)^2 + \sum_{a=4}^6\left(
\frac{\mu}{6}\right)^2 (x^a)^2
\right)(dx^+)^2 + \sum_{I=1}^9(dx^{I})^2\,, \\
F_{+123} &=& \mu\,. \nn
\end{eqnarray}
The action of the matrix model on this background $S_{\rm pp}$ consists
of two parts as follows:\footnote{Hereafter we will rescale the gauge
field and parameters as $ A \rightarrow R A\,,~ t \rightarrow t/R\,,~
\mu \rightarrow R \mu$\,.}
\begin{align}
S_{\rm pp} & = S_\mathrm{flat} + S_\mu ~, \\ S_\mathrm{flat} & = \int\!\! 
dt\, \mathrm{Tr} \left[ \frac{1}{2R} D_t X^I D_t X^I + \frac{R}{4} ( [
X^I, X^J] )^2 + i \Theta^\dagger D_t \Theta - R \Theta^\dagger \gamma^I
[ \Theta, X^I ] \right] ~, \label{o-action} \\ S_\mu &= \int\!\! dt\,
\mathrm{Tr} \left[ -\frac{1}{2R} \left( \frac{\mu}{3} \right)^2 (X^i)^2
-\frac{1}{2R} \left( \frac{\mu}{6} \right)^2 (X^a)^2 - i \frac{\mu}{3}
\epsilon^{ijk} X^i X^j X^k - i \frac{\mu}{4} \Theta^\dagger \gamma^{123}
\Theta \right] ~\,, \notag
\end{align}
where the indices of the transverse nine-dimensional space are 
$I,J=1,\ldots,9$ and $R$ is the radius of the circle compactified
along $x^-$\,. All degrees of freedom are $N\times N$ Hermitian matrices
and the covariant derivative $D_t$ with the gauge field $A$ is defined
by $D_t = \partial_t-i[A,~~]$\,. This matrix model is closely related to
the supermembrane theory on the pp-wave background via the matrix
regularization \cite{dWHN} (For works in this direction see \cite{DSR,SY}). 
This matrix model has a supersymmetric fuzzy sphere solution 
(which is called ``giant
graviton'') due to the Myers effects \cite{Myers}, because the constant
4-form flux is equipped with. 
This fuzzy sphere solution is given by
\begin{equation}
X^i_\mathrm{sphere} = \frac{\mu}{3} J^i ~,
\label{fuzzy}
\end{equation}
where $J^i$ satisfies the $SU(2)$ algebra $[ J^i, J^j ] = i
\epsilon^{ijk} J^k$\,. The classical energy of this solution is zero and
hence the fuzzy sphere can appear in classical vacua without loss of
energy. Namely, the classical vacua of the pp-wave matrix model are
enriched with fuzzy spheres, and it may be interesting to look deeper
into the dynamics of fuzzy sphere solution. Then the fuzzy sphere
solution $X^i_\mathrm{sphere}$ preserves the full 16 dynamical
supersymmetries of the pp-wave and hence is 1/2-BPS object.  
We note that actually there is another fuzzy sphere solution of the form
$\frac{\mu}{6} J^i$ but such solution does
not have quantum stability and is thus non-BPS object \cite{SY3}  
(For other classical solutions, see \cite{Bak,HS1,Park,sol}). 
In \cite{HSKY-potential} the interaction potential of membrane fuzzy spheres 
was shown to be the $1/r^7$-type.

\begin{figure}
 \begin{center}
  \includegraphics[scale=0.6]{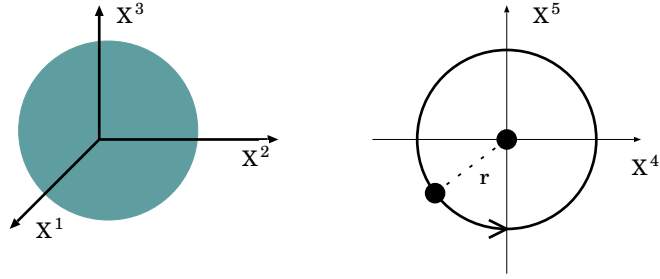}
\caption{\footnotesize The two-body system of fuzzy spheres. A fuzzy
  sphere rotates around the other one sitting at the origin in the
  transverse six-dimensional space.}
\label{flat:fig} 
 \end{center}
\end{figure}

In this letter we discuss a thermal interaction between fuzzy membrane
solutions in the pp-wave matrix model. The two-body interaction of fuzzy
spheres at zero temperature was studied in our previous work \cite{HSKY}
(containing the extension of the work \cite{Huang}) by considering the
system that a fuzzy sphere rotates with a constant radius $r$ around the
other one sitting at the origin in the transverse six-dimensional space
(Fig.\,\ref{flat:fig}). In the case of zero temperature the interaction
potential is zero basically because of the existence of the remaining
supersymmetries. In the finite temperature case, however, non-trivial
interaction should appear since the supersymmetries are completely
broken due to the thermal effect. Here, we will be interested in the
evaluation of the thermal potential at finite temperature by heating the
system depicted in Fig.\,\ref{flat:fig}. Firstly we consider the exact
one-loop free energy by using the $\mu\rightarrow\infty$ limit. In this
computation we use the spectrum around the two-body system which has
been obtained in \cite{HSKY}. By numerically plotting the free energy,
we can see that the rotating fuzzy sphere tends to approach to the
static fuzzy sphere. The attractive potential becomes stronger and
stronger as temperature grows and the distance decreases. Furthermore,
as the size of the matrix becomes larger and larger, the potential
becomes deeper and deeper.
In addition we evaluate the analytical expression of the free energy by
taking the large $N$ limit. Then the discrete sum for the index of spin
may be replaced by the integral for it. When we utilize the
long-distance expansion, we find that the leading part of the free
energy can be represented by a polylog function. The approximation of
the free energy qualitatively well reproduces the behavior of the exact
free energy.

This paper is organized as follows: In section2 we introduce the formula
for the exact free energy of a pp-wave matrix model at finite
temperature. In section 3 the free energy for the interaction between
two fuzzy spheres is evaluated both in numerical and analytical
methods.

\section{One-Loop Exact Free Energy at Finite Temperature} 

We will briefly review the calculus of the one-loop free energy in
the finite temperature case.  By using the background field method and
the $\mu\rightarrow\infty$ limit, it can
be exactly computed. 

We first decompose the variables as 
\begin{equation}
X^I = B^I + Y^I\,, \qquad \Theta = 0 + \Psi\,,
\label{cl+qu}
\end{equation}
where $B^I$ are the classical background fields while $Y^I$ and
$\Psi$ are the quantum fluctuations around them. 
In some cases we need to consider the background of the gauge field
(gauge field moduli) due to the finite temperature effect.  In the case
we consider here, however, we do not need to take account of it, as we
will explain later.

In order to perform the path integration, we take the background field
gauge which is usually chosen in the matrix model calculation as
\begin{equation}
D_\mu^{\rm bg} A^\mu_{\rm qu} \equiv 
D_t A + i [ B^I, X^I ] = 0 ~.
\label{bg-gauge}
\end{equation}
Then the corresponding gauge-fixing $S_\mathrm{GF}$ and Faddeev-Popov
ghost $S_\mathrm{FP}$ terms are given by
\begin{equation}
S_\mathrm{GF} + S_\mathrm{FP} = \int\!dt \,{\rm Tr} \left( - \frac{1}{2}
(D_\mu^{\rm bg} A^\mu_{\rm qu} )^2 - \bar{C} \partial_{t} D_t C + [B^I,
\bar{C}] [X^I,\,C] \right) ~.  \label{gf-fp}
\end{equation}
Now by inserting the decomposition of the matrix fields (\ref{cl+qu})
into the matrix model action, we get the gauge fixed plane-wave action
$S$ $(\equiv S_{\rm pp} + S_\mathrm{GF} + S_\mathrm{FP})$ expanded around
the background.  The resulting acting is read as 
$S =  S_0 + S_2 + S_3 + S_4$\,,  
where $S_n$ represents the action of order $n$ with respect to the
quantum fluctuations and, for each $n$, its expression is
\begin{align}
S_0 = \int dt \, \mathrm{Tr} \bigg[ \,
&      \frac{1}{2}(\dot{B}^I)^2  
        - \frac{1}{2} \left(\frac{\mu}{3}\right)^2 (B^i)^2 
        - \frac{1}{2} \left(\frac{\mu}{6}\right)^2 (B^a)^2 
        + \frac{1}{4}([B^I,\,B^J])^2
        - i \frac{\mu}{3} \epsilon^{ijk} B^i B^j B^k 
    \bigg] ~,
\notag \\
S_2 = \int dt \, \mathrm{Tr} \bigg[ \,
&       \frac{1}{2} ( \dot{Y}^I)^2 - 2i \dot{B}^I [A, \, Y^I] 
        + \frac{1}{2}([B^I , \, Y^J])^2 
        + [B^I , \, B^J] [Y^I , \, Y^J]
        - i \mu \epsilon^{ijk} B^i Y^j Y^k
\notag \\
&       - \frac{1}{2} \left( \frac{\mu}{3} \right)^2 (Y^i)^2 
        - \frac{1}{2} \left( \frac{\mu}{6} \right)^2 (Y^a)^2 
        + i \Psi^\dagger \dot{\Psi} 
        -  \Psi^\dagger \gamma^I [ \Psi , \, B^I ] 
        -i \frac{\mu}{4} \Psi^\dagger \gamma^{123} \Psi  
\notag \\ 
&       - \frac{1}{2} \dot{A}^2  - \frac{1}{2} ( [B^I , \, A])^2 
        + \dot{\bar{C}} \dot{C} 
        + [B^I , \, \bar{C} ] [ B^I ,\, C] \,
     \bigg] ~,
\notag \\
S_3 = \int dt \, \mathrm{Tr} \bigg[
&       - i\dot{Y}^I [ A , \, Y^I ] - [A , \, B^I] [ A, \, Y^I] 
        + [ B^I , \, Y^J] [Y^I , \, Y^J] 
        +  \Psi^\dagger [A , \, \Psi] 
\notag \\
&       -  \Psi^\dagger \gamma^I [ \Psi , \, Y^I ] 
        - i \frac{\mu}{3} \epsilon^{ijk} Y^i Y^j Y^k
        - i \dot{\bar{C}} [A , \, C] 
        +  [B^I,\, \bar{C} ] [Y^I,\,C]  \,
     \bigg] ~,
\notag \\
S_4 = \int dt \, \mathrm{Tr} \bigg[
&       - \frac{1}{2} ([A,\,Y^I])^2 + \frac{1}{4} ([Y^I,\,Y^J])^2 
     \bigg] ~.
\label{bgaction} 
\end{align}

For the justification of one-loop computation or the semi-classical
analysis, it should be made clear that $S_3$ and $S_4$ of
Eq.~(\ref{bgaction}) can be regarded as perturbations.  For this
purpose, following \cite{DSR}, we rescale the fluctuations and
parameters as
\begin{gather}
A   \rightarrow \mu^{-1/2} A   ~,~~~
Y^I \rightarrow \mu^{-1/2} Y^I ~,~~~
C  \rightarrow \mu^{-1/2} C   ~,~~~
\bar{C} \rightarrow \mu^{-1/2} \bar{C} ~,~~~ 
t \rightarrow \mu^{-1} t ~.
\label{rescale}
\end{gather}
Under this rescaling, the action $S$ in the fuzzy sphere background becomes
\begin{align}
S =  S_2 + \mu^{-3/2} S_3 + \mu^{-3} S_4 ~,
\label{ssss}
\end{align}
where $S_2$, $S_3$ and $S_4$ do not have $\mu$ dependence.  Now it is
obvious that, in the large $\mu$ limit, $S_3$ and $S_4$ can be treated
as perturbations and the one-loop computation gives the sensible
result. Note that the analysis in the $S_2$ part is exact in the
$\mu\rightarrow \infty$ limit. 
We can calculate the exact spectra around an $N$-dimensional
irreducible fuzzy sphere in the $\mu \rightarrow \infty$ limit, by
following the method in the work \cite{DSR} (For the detail
calculation, see \cite{DSR,HSKY}).  

Now we are interested in the finite temperature case where the system
couples to the thermal bath. In order to consider the thermal system
with temperature $T$, let us move to the Euclidean formulation via the
Wick rotation $t\rightarrow it$, and compactify the Euclidean time
direction with a periodicity $\beta \equiv 1/T$.  Note that $T$ is a
dimensionless parameter now because of the scaling of time variable $t
\rightarrow R^{-1}t$.  This compactification leads us to encounter the 
summation instead of the momentum integral. 
We can easily compute this summation by using the formulae 
\begin{eqnarray}
\sum_{n={\rm integer}}\ln\left(n^2\pi^2 + M'{}^2\right) = 2\ln\sinh M' \quad
 (\mbox{for bosons})\,,
 \\
\sum_{n={\rm half~integer}}\ln\left(n^2\pi^2 + M'{}^2\right) = 2\ln\cosh M'
 \quad (\mbox{for fermions})\,, 
\end{eqnarray} 
and the fact that the fuzzy sphere configuration under our consideration
is supersymmetric. Here we should note that the ghost fields obey the
periodic condition rather than anti-periodic condition. Hence its
contribution gives the sinh but the integration of the ghost field 
gives the inverse sign in comparison to bosons. This fact ensures the
cancellation among unphysical and ghost degrees of freedom.  
Thus the free energy $F = - T \ln Z$ is
represented by
\begin{eqnarray}
F = T\sum_{i\in Y,A}N_i \ln\left(1 - \e^{-\beta M_i}\right) 
- \frac{1}{2}T N_{\Psi}
\ln\left(1 + \e^{-\beta M_{\Psi}}\right) - 2T N_C 
\ln\left(1 - \e^{-\beta M_C} \right)\,. \label{general}
\end{eqnarray}
The symbols $Y$ and $A$ denote the bosonic
fluctuations of $X$ and $A$, respectively. The $\Psi$ and $C$ denote the
fermionic fluctuations of $\Theta$ and $C$\,. The free energy usually
contains the zero temperature part, but this part does not appear in the
present case since the fuzzy sphere background is supersymmetric at zero
temperature.

\section{Free Energy of Fuzzy Membrane Interaction} 

We are interested in the configuration of classical solution in which
a fuzzy sphere rotates with a constant radius $r$ around the other fuzzy
sphere (see Fig.\,\ref{flat:fig}). This system is described by 
\begin{eqnarray}
&& B^{I} = \begin{pmatrix}
B_{(1)}^I & 0 \\ 0 & B_{(2)}^I 
\end{pmatrix}\,, \quad  B^i_{(s)}=\frac{\mu}{3}J^i_{(s)} \quad (i=1,2,3~;~s=1,2)\,, 
\nn \\ 
&& 
B_{(1)}^4 = r\cos\left(\frac{\mu}{6}t\right)
{\bf 1}_{N_1\times N_1}\,, \quad B^5_{(1)} = r\sin\left(\frac{\mu}{6}t\right)
{\bf 1}_{N_1\times N_1}\,, \quad \mbox{otherwise}=0\,.  
\label{clfzs}
\end{eqnarray}
The $B_{(s)}^{I}$ are $N_s\times N_s$ matrices. We take $B^I$ as
$N\times N$ matrices and then $N=N_1+N_2$\,. Let us concentrate on the
$N_1\neq N_2$ case so that we do not have to take account of the gauge
field moduli considered in \cite{Furuuchi,Semenoff,Semenoff2}.  
If two fuzzy spheres of the
same size are considered then we may have the gauge field moduli. 
This gauge field moduli plays an important role in the study of the 
Hagedorn transition in the transverse M5-brane vacua \cite{TM5}. 

This system (\ref{clfzs}) was proposed in our previous work \cite{HSKY}
and it was shown to be supersymmetric. That is, the interaction
potential between them vanishes (one-loop flatness) \cite{HSKY} as well
as the quantum fluctuations around each of the fuzzy spheres
\cite{DSR,SY3,HSKY}. This result comes from the supersymmetries of the
theory. However, the supersymmetries are completely broken down and so
non-trivial potential should appear. In our previous work
\cite{HSKY-thermal} we studied the stability of the fuzzy
sphere
\footnote{Stability of fuzzy spheres is discussed in the IIB
matrix model with Chern-Simons term\cite{IIB-CS} and mass term
\cite{massiveIIB}.} 
in the case of finite temperature. In particular, we
compared the free energies between the fuzzy sphere vacuum $(X^i=
\frac{\mu}{3}J^i~(i=1,2,3))$ and the trivial one $(X^I=0)$\,, where all
the fuzzy spheres are considered to be located at the origin in the
transverse six-dimensional space. Then we are now interested in the
interaction potential between the two fuzzy spheres being at a distance
in the transverse space and will consider the off-diagonal fluctuations
which are responsible for the interaction.

\begin{table}
{\small 
\begin{center}
 \begin{tabular}{lccc}
\hline
Fields  & (Mass)$^2$ & Degeneracy & Spin \\
\hline 
\underline{$SO(3)$ bosons}  & & & \\
$\al_{jm}$ & $r^2+\frac{1}{3^2}(j+1)^2$ & $2j+1$ & $\frac{1}{2}|N_1-N_2| 
\leq j \leq \frac{1}{2}(N_1+N_2)-2$ \\
$\beta_{jm}$ & $r^2+\frac{1}{3^2}j^2$ & $2j+1$ & $\frac{1}{2}|N_1-N_2|+1 
\leq j \leq \frac{1}{2}(N_1+N_2)$ \\  
$\omega_{jm}$ ~(gauge) &  $r^2+\frac{1}{3^2}j(j+1)$ & $2j+1$ & 
$\frac{1}{2}|N_1-N_2| \leq j \leq \frac{1}{2}(N_1+N_2)-1$ \\ 
\underline{$SO(4)$ bosons} & & & \\  
$\phi_{jm}^{a'}~(a'=6,7,8,9)$ 
& $r^2+\frac{1}{3^2}\left(j+\frac{1}{2}\right)^2$ & $4(2j+1)$ & 
$\frac{1}{2}|N_1-N_2|\leq j \leq \frac{1}{2}(N_1+N_2)-1$ \\ 
\underline{Rotational Part} & & & \\ 
$(r1)_{jm}$~(gauge) & $r^2+\frac{1}{3^2}j(j+1)$ & $2j+1$ 
& $\frac{1}{2}|N_1-N_2|\leq j \leq 
\frac{1}{2}(N_1+N_2)-1$ \\ 
$(r2)_{jm}$  
& $r^2+\frac{1}{3^2}j^2$ & $2j+1$ & $\frac{1}{2}|N_1-N_2|\leq j \leq 
\frac{1}{2}(N_1+N_2)-1$ \\ 
$(r3)_{jm}$ 
& $r^2+\frac{1}{3^2}(j+1)^2$ & $2j+1$ & $\frac{1}{2}|N_1-N_2|\leq j \leq 
\frac{1}{2}(N_1+N_2)-1$ \\ 
\underline{Fermion 1} & & & \\ 
$(F_{1})_{jm}$ & $r^2+\frac{1}{3^2}(j+1)^2$ & $2(2j+1)$ &
  $\frac{1}{2}|N_1-N_2|-\frac{1}{2} \leq j \leq \frac{1}{2}(N_1+N_2)
-\frac{3}{2}$ \\ 
$(F_{2})_{jm}$ 
& $r^2+\frac{1}{3^2}\left(j+\frac{1}{2}\right)^2$ & $2(2j+1)$ & 
  $\frac{1}{2}|N_1-N_2|-\frac{1}{2} \leq j \leq \frac{1}{2}(N_1+N_2)
-\frac{3}{2}$ \\ 
\underline{Fermion 2} & & & \\ 
$(F_1')_{jm}$ & $r^2+\frac{1}{3^2}\left(j+\frac{1}{2}\right)^2$ & $2(2j+1)$ & 
$\frac{1}{2}|N_1-N_2|+\frac{1}{2}\leq j \leq \frac{1}{2}(N_1+N_2)
-\frac{1}{2}$ \\ 
$(F_2')_{jm}$ &  $r^2+\frac{1}{3^2}j^2$ & $2(2j+1)$ & 
$\frac{1}{2}|N_1-N_2|+\frac{1}{2}\leq j \leq \frac{1}{2}(N_1+N_2)
-\frac{1}{2}$ \\ 
\underline{Ghost} & & & \\ 
$c_{jm}~(\bar{c}_{jm})$ & $r^2+\frac{1}{3^2}j(j+1)$ & $2j+1$ & $\frac{1}{2}|N_1-N_2|\leq j \leq 
\frac{1}{2}(N_1+N_2)-1$ \\ \hline  
 \end{tabular}
\caption{\footnotesize The spectrum of the interaction part between two
fuzzy spheres.} \label{spectrum:tab}
\end{center}
} 
\end{table} 

In the computation of the partition function for off-diagonal
fluctuations, we should diagonalize the quadratic action. After
the diagonalization, we see the spectrum. The resulting
spectrum is summarized in Tab.\,\ref{spectrum:tab}.  By using the
spectrum the free energy is expressed as
\begin{eqnarray}
\frac{1}{2}\beta F &=& \sum_{j=\frac{1}{2}|N_1-N_2|}^{\frac{1}{2}(N_1+N_2)-2}
\!\!\!\!(2j+1)\ln(1-\e^{-\beta\sqrt{r^2+\frac{1}{3^2}(j+1)^2}}) 
+ \sum_{j=\frac{1}{2}|N_1-N_2|+1}^{\frac{1}{2}(N_1+N_2)}\!\!\!\! 
(2j+1)\ln(1-\e^{-\beta\sqrt{r^2+\frac{1}{3^2}j^2}}) \nn \\
&& 
+ \sum_{j=\frac{1}{2}|N_1-N_2|}^{\frac{1}{2}(N_1+N_2)-1}\!\!\!\!
4(2j+1)\ln(1-\e^{-\beta\sqrt{r^2+\frac{1}{3^2}\left(j+\frac{1}{2}\right)^2}}) 
\nn \\ && 
+ \sum_{j=\frac{1}{2}|N_1-N_2|}^{\frac{1}{2}(N_1+N_2)-1}\!\!\!\!(2j+1)
\left[
\ln(1-\e^{-\beta\sqrt{r^2+\frac{1}{3^2}j^2}}) 
+ \ln(1-\e^{-\beta\sqrt{r^2+\frac{1}{3^2}(j+1)^2}})  
\right] \nn \\
&& - \sum_{j=\frac{1}{2}|N_1-N_2|-\frac{1}{2}}^{\frac{1}{2}(N_1+N_2)
-\frac{3}{2}}\!\!\!\!2(2j+1)\left[
\ln(1+\e^{-\beta\sqrt{r^2+\frac{1}{3^2}(j+1)^2}}) + 
\ln(1+\e^{-\beta\sqrt{r^2+\frac{1}{3^2}(j+\frac{1}{2})^2})} 
\right] \nn \\ 
&& - \sum_{j=\frac{1}{2}|N_1-N_2|+\frac{1}{2}}^{\frac{1}{2}(N_1+N_2)
-\frac{1}{2}}\!\!\!\!2(2j+1)\left[
\ln(1+\e^{-\beta\sqrt{r^2+\frac{1}{3^2}(j+\frac{1}{2})^2}}) + 
\ln(1+\e^{-\beta\sqrt{r^2+\frac{1}{3^2}j^2}}) 
\right] 
\,. 
\end{eqnarray}
Here the unphysical degrees of freedom are canceled out. 
That is, the above expression has no degrees of freedom for fuzzy sphere
rotation, gauge field and ghosts.

We will consider the free energy of the fuzzy membrane interaction in
both numerical and analytical methods below. To begin with, we
numerically study the behavior of the free energy. Then we evaluate the
analytical expression of the free energy by using the large $N$ limit.

\subsection{Numerical Study of the Free Energy} 

\begin{figure}
 \begin{center}
\begin{tabular}{lc}
  \includegraphics[scale=.6]{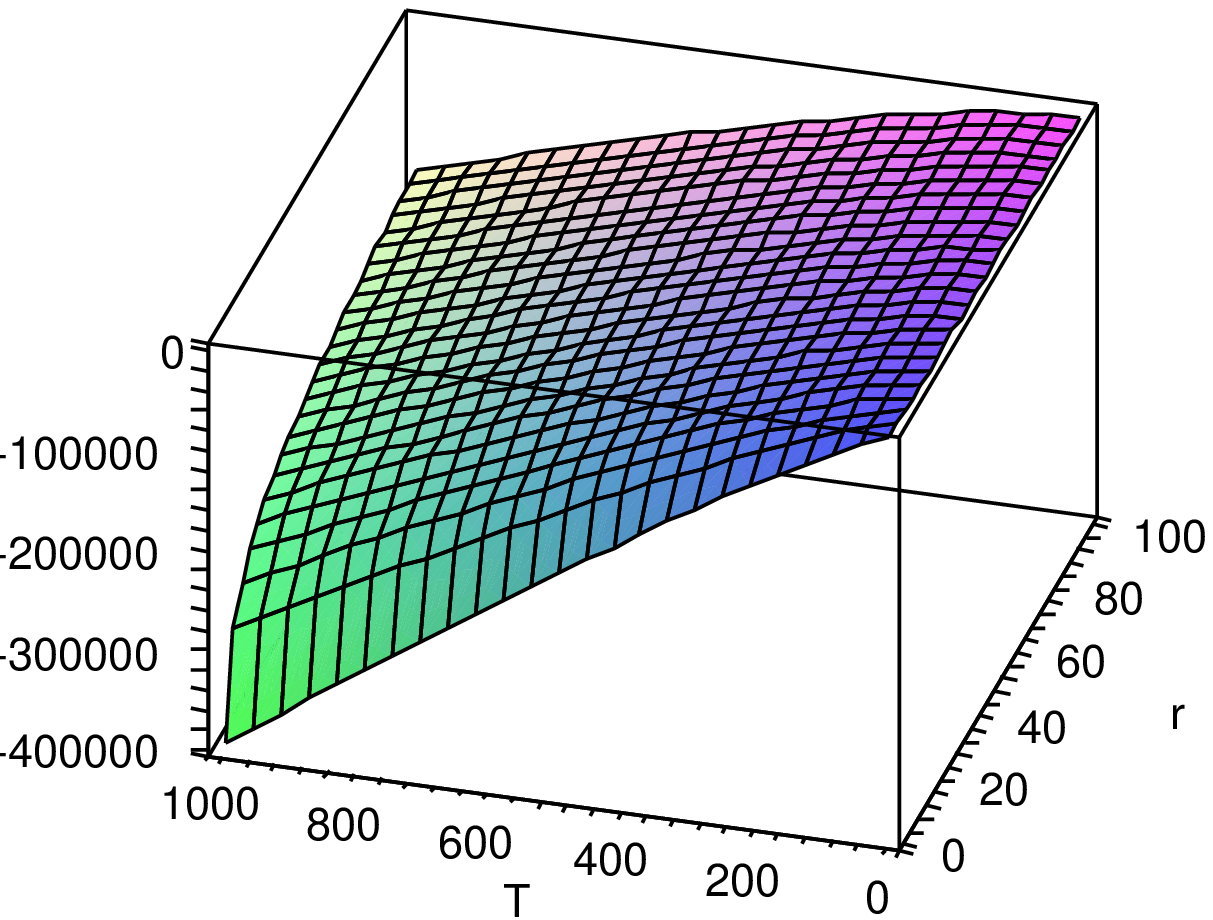}  &  
  \includegraphics[scale=.6]{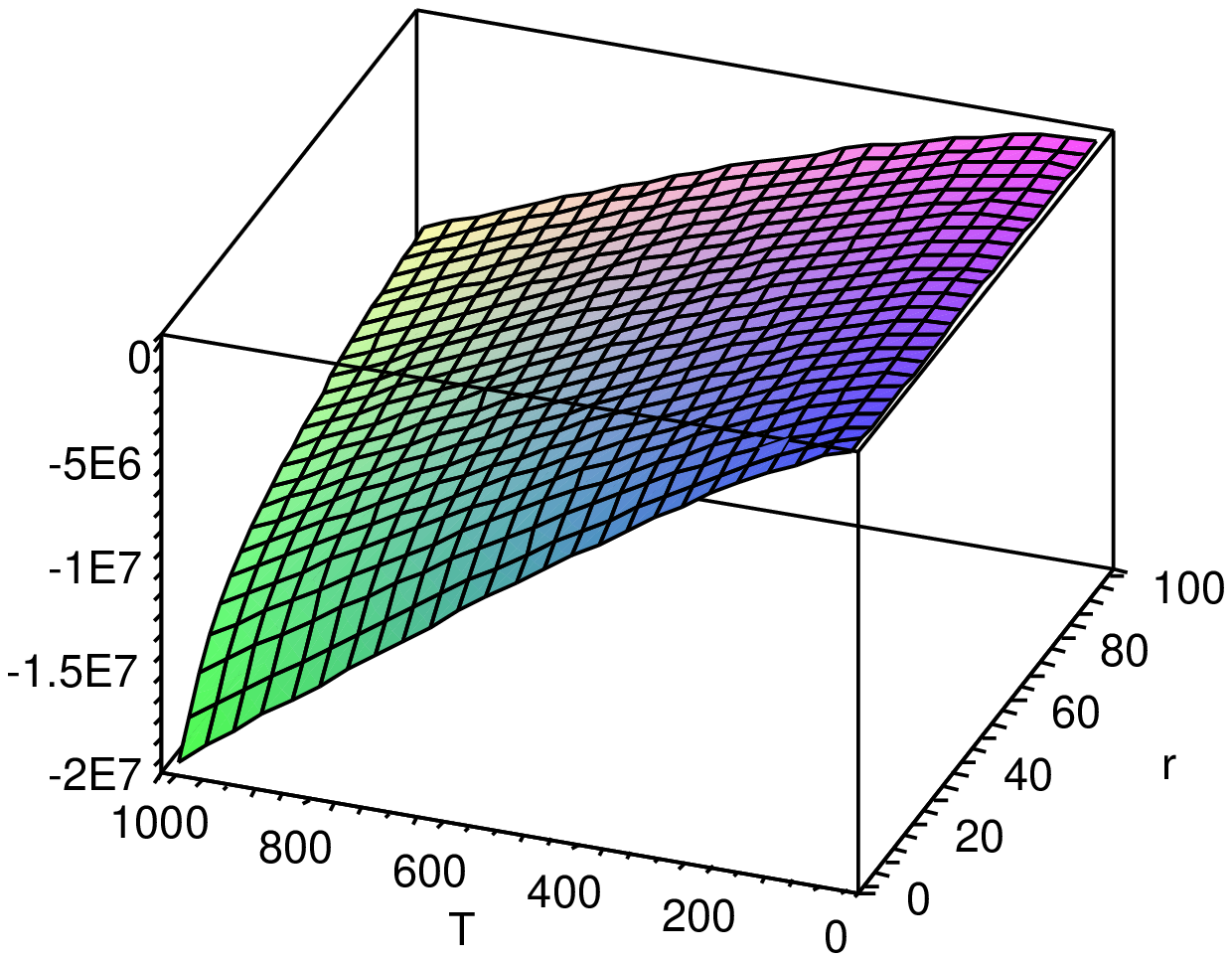}  \\ 
\qquad\qquad\quad {\footnotesize i)~~$N_1=3$ and $N_2=2$} & 
{\footnotesize ii)~~$N_1=20$ and $N_2=10$}
\end{tabular}
\caption{\footnotesize The numerical plots of the free energy $F$ for
the two-body interaction. The $F$ is plotted in the vertical axis with
respect to the distance $r$ and the temperature $T$\,. The left one is
for the $N_1=3$ and $N_2=2$\,, and the right is for $N_1=20$ and
$N_2=10$\,.}  \label{free:fig}
 \end{center}
\end{figure}

The expression of the free energy for the fuzzy membrane interaction is
quite complicated and so it is difficult to study analytically the
behavior of the free energy without any approximation. We will analyze
the analytic behavior of the free energy in the next subsection by using
the large $N$ limit and the long-range expansion. Here let us however
investigate numerically the behavior of the free energy. In
Fig.\,\ref{free:fig}, we give the numerical plot of the free energy for
some cases. From the results we can see that the interaction between
two fuzzy spheres is attractive.

\subsection{Analytic Behavior of the Free Energy in Large $N$ Limit} 

\begin{figure}
 \begin{center}
  \includegraphics[scale=.5]{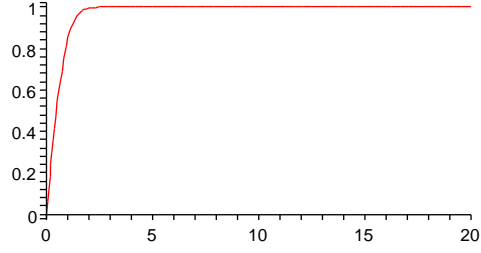} 
\caption{\footnotesize The error function. }
\label{erf:fig}
 \end{center} 
\end{figure}

\begin{figure}
 \begin{center} 
\begin{tabular}{cc} 
  \includegraphics[scale=.6]{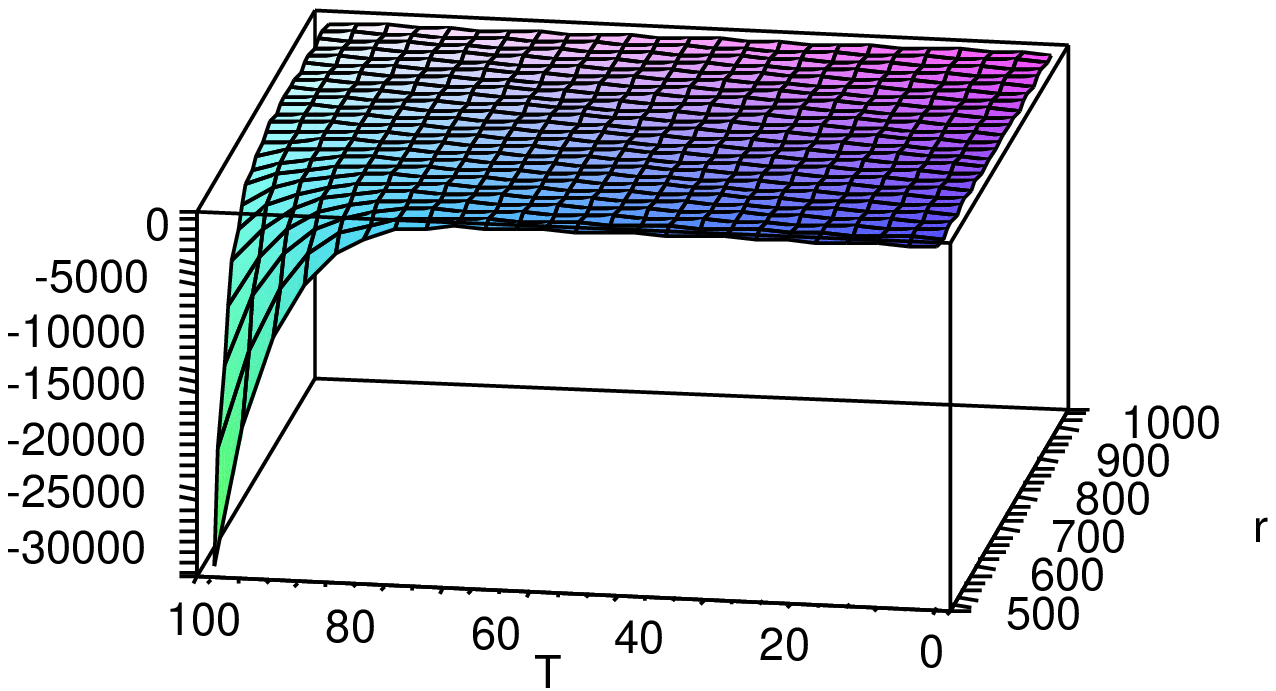} & 
\includegraphics[scale=.6]{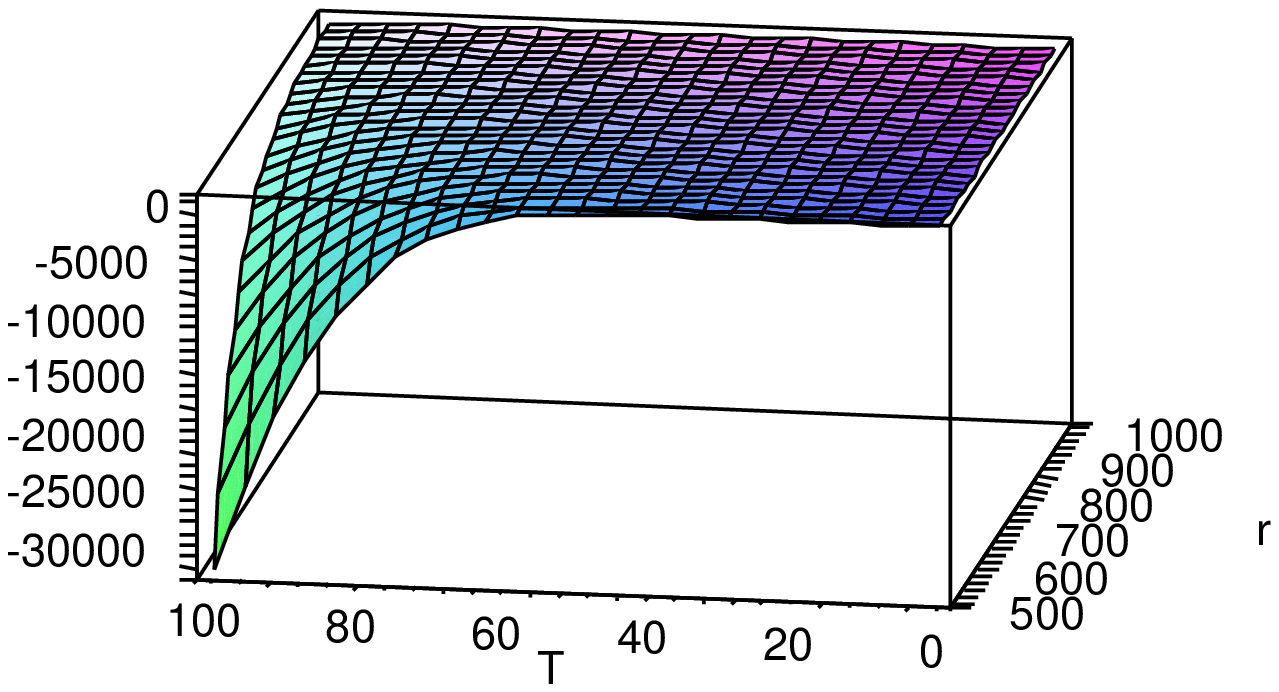}  \\ 
{\footnotesize i)~~the approximated free energy in large $N$} 
& {\footnotesize ii)~~the exact free energy for $N_1=50$ and
 $N_2=30$} 
\end{tabular}
  \caption{\footnotesize A comparison of the approximated free energy in
  large $N$ and the exact free energy. In i) the approximated free
  energy is numerically plotted. In ii) the exact free energy in the
  same region as i) is shown in order to see the validity of the
  approximation.}  \label{largeN:fig}
 \end{center}
\end{figure}

In the previous subsection we presented the numerical study of the
thermodynamic behavior of the free energy for the interaction between
two fuzzy spheres. From now on we will evaluate the analytical
expression by considering the large $N$ limit. In such situations we can
replace the discrete summation for the spin $j$ by an integral with a
continuum variable and so it is possible to evaluate an analytical
expression for the free energy. For example, let us focus upon a part of
the $SO(3)$ bosons, $\al_{jm}$\,. We consider the large $N$ limit:
\begin{eqnarray}
N_1,~N_2 \gg 1 \quad \mbox{with} \quad N_1-N_2 =c\,, 
\end{eqnarray}
where $c$ is a non-vanishing constant. 
Then the summation for $\al_{jm}$ may be replaced as follows: 
\begin{eqnarray}
\sum_{j=\frac{1}{2}|N_1-N_2|}^{\frac{1}{2}(N_1+N_2)-2}\!\!(2j+1)\ln\left(
1-\e^{-\beta\sqrt{r^2+\frac{1}{3^2}(j+1)^2}}
\right) \longrightarrow 
\int^N_{\frac{1}{2}|c|} \!\!dx\,(2x+1)\ln(1-\e^{-\beta\sqrt{r^2+\frac{1}{3^2}(x+1)^2}})\,. 
\end{eqnarray}
In addition, in order to evaluate the above integral analytically, we
utilize the long-distance expansion (large $r$ limit) and expand the
square root as
\begin{eqnarray}
\sqrt{r^2+\frac{1}{3^2}(x+1)^2} \cong r + \frac{1}{18r}(x+1)^2 + \mathcal{O}\left(\frac{1}{r^3}\right)\,. 
\end{eqnarray} 
By using the Taylor expansion,
\begin{eqnarray}
\ln(1-y) = -\left(y + \frac{1}{2}y^2 + \frac{1}{3}y^3 + \cdots \right)\,, 
\end{eqnarray} 
the free energy for $\al_{jm}$ can be evaluated as 
\begin{eqnarray}
F_{\al} &=& 36T^2 r \sum_{n=1}^{\infty} \frac{1}{n^2} \left\{
\e^{-\frac{1}{18}\left(\frac{n\beta}{r}\right)(18r^2+1+2N+N^2)} 
- \e^{-\frac{1}{72}\left(\frac{n\beta}{r}\right)(72r^2+4+4|c|+|c|^2)} 
\right\} \nn  \\ 
&& +3\sqrt{2\pi r}T^{3/2}\sum_{n=1}^{\infty} \left\{
\mbox{erf}\left(\frac{N+1}{6}\sqrt{\frac{2n\beta}{r}} \right) 
-\mbox{erf}\left(\frac{|c|+2}{12}\sqrt{\frac{2n\beta}{r}}\right)
\right\}
\frac{\e^{-n\beta r}}{n^{3/2}} 
\nn \\
&\cong&  -36T^2r \sum_{n=1}^{\infty}\frac{1}{n^2}
\e^{- n\beta r} + 3\sqrt{2\pi r}T^{3/2}\sum_{n=1}^{\infty}\left\{
1-\mbox{erf}\left(\frac{|c|+2}{12}\sqrt{\frac{2n\beta}{r}}\right)\right\}
\frac{\e^{-n\beta r}}{n^{3/2}} 
\,, 
\label{pre}
\end{eqnarray}
where we have used that $N$ and $r$ is sufficiently large 
so that erf$(x)\cong 1~~(x\gg 1)$ as shown in Fig.\,\ref{erf:fig}, 
but we have supposed that $r\gg N$\,. 
When we consider the low temperature region $T \ll 1/r$\,, the second
summation in the expression (\ref{pre}) can be ignored. 
Then, using a polylog function,
\[
 \mbox{polylog}(a,z) \equiv \sum_{n=1}^{\infty}\frac{z^n}{n^a}\,. 
\] 
we can write down the analytical expression of the free energy as
follows: 
\begin{eqnarray}
\label{boson}
F \cong -36T^2 r \cdot \mbox{polylog}(2,\e^{-\beta r})\,. 
\end{eqnarray}
For the other bosonic sector we can evaluate the free energy in the
similar way and the same result as (\ref{boson})\,. Then the fermionic
sector and the ghost sector the contributions are evaluated as
 \begin{eqnarray}
\label{fermion}
F \cong -36T^2 r \cdot\mbox{polylog}(2,-\e^{-\beta r})\,. 
\end{eqnarray} 
Thus the total free energy is given by 
\begin{eqnarray}
F &\sim& - 36 \times 8 T^2 r \Bigl[
{\rm polylog}(2,\e^{-\beta r}) + {\rm polylog}(2, -\e^{-\beta r})
\Bigr] \nn \\ 
&=& -144 T^2 r\cdot {\rm polylog}(2,\e^{-2\beta r})\,,
\label{largeN}
\end{eqnarray}
in the region:
\begin{eqnarray}
T \ll \frac{1}{r}\,, \qquad N \ll r\,, \qquad 
r,~N:~\mbox{sufficiently large}\,. 
\end{eqnarray}
Here we have used the formula: 
\[
 {\rm polylog}(2,z) + {\rm polylog}(2, -z) = 
\frac{1}{2}{\rm polylog}(2,z^2)\,. 
\]
In particular, polylog$(2,1)=\pi^2/6$ and polylog$(2,0)=0$\,. 
The numerical plot of (\ref{largeN}) is shown in
Fig.\,\ref{largeN:fig}. We see that the behavior of (\ref{largeN}) 
is quantitatively similar to that of the exact free energy with large
$N$ values. In the low temperature region the large $N$ free energy for
the membrane fuzzy sphere interaction is well described by a polylog
function. 

Note that we cannot naively evaluate the high-temperature behavior 
from (\ref{pre}) since the $n\cong\infty$ contributes and so we cannot
take the error function as erf(0)=0.

\section*{Acknowledgments} 

One of us, K.Y., would like to thank N.~Kawahara, J.~Nishimura,
Y.~Takayama and D.~Tomino for helpful discussions. 
The work of H.S. was supported by grant No. R01-2004-000-10651-0
from the Basic Research Program of the Korea Science and
Engineering Foundation (KOSEF). 
The work of K.~Y.\ is supported in part by JSPS
Research Fellowships for Young Scientists.

\end{document}